\documentclass[pra,aps,showpacs,twocolumn,floatfix]{revtex4}
\usepackage{graphicx}
\usepackage[ansinew]{inputenc}
\usepackage{array}
\usepackage{color}
\usepackage{amsmath}
\usepackage{amsxtra}
\usepackage{amstext}
\usepackage{amssymb}
\usepackage{latexsym}
\usepackage{dsfont}
\usepackage{verbatim}
\usepackage{comment}

\begin{document}

\title
{Metal nanoparticle plasmons operating within quantum lifetime }

\author{Mehmet Emre Ta\c{s}g\i n}
\affiliation{Department of Electrical \& Electronics Engineering, K{\i}rklareli University, 39020 Karah{\i}d{\i}r, K{\i}rklareli, Turkey}

\date{\today}

\begin{abstract}
We investigate the dynamics of a plasmonic oscillation over a metal nanoparticle when it is strongly coupled to a quantum emitter (e.g. quantum dot, molecule). We simulate the density matrix evolution for a simple model; coupled classical--quantum oscillators system. We show that lifetime of the plasmonic oscillations can be increased several orders of magnitude, upto the decay time of the quantum emitter. This effect shows itself as the narrowing of the plasmon emission band in the spaser (surface plasmon amplification by stimulated emission of radiation) experiment [{\it Nature}, {\bf 2009}, 460, 1110], where a gold nanoparticle interacts with the surrounding molecules.  Enhancement of the plasmonic excitation lifetime enables stimulated emission to overcome the spontaneous one. The enhancement occurs due to the emergence of a phenomenon analogous to electromagnetically induced transparency (EIT). The effect can find applications in many areas of nanoscale physics, such as in quantum information with plasmons and in increasing solar cell efficiency.
\end{abstract}


\maketitle

\section{Introduction}





Ordinarily, the response of atoms to a electromagnetic field mimics a two-level quantum system since the degeneracy in the excited levels is masked out by dipole selection rules. However, if the dipole allowed excited state is microwave coupled to the forbidden one, the dielectric response will be severely modified \cite{Scullybook}. A transparency window emerges at the resonance frequency, where without microwave drive an absorption peak would have been observed. This phenomenon is called as Electromagnetically Induced Transparency (EIT). The vanishing absorption is due to the following. The microwave coupling splits the dipole allowed excited state into two (Stark effect \cite{stark}). Since the splitting is smaller than the decay rate of the excited level, optical beam couples the ground state to each auxiliary levels. When the two Rabi oscillations are out of phase, absorption cancels and atom stays in the ground state \cite{Harris,Alzar,Alzar2,Tokman}. The modified absorption is proportional to the decay rate of the dipole forbidden transition (much smaller) instead of the allowed one \cite{Scullybook}. EIT-like schemes underlie in the physics of ultra-slow light propagation \cite{SlowLight,SlowYanik} and index enhancement \cite{Fleischhauer}.

A similar phenomenon shown in Ref.s\cite{Alzar,Alzar2,Tokman} takes place for the two coupled classical oscillators (C.O.), where the first one has a high damping rate (low-quality) and the second one has a low damping rate (high-quality). When the low-quality C.O. is driven by a harmonic force, the absorbed power is governed with the damping rate of the high-quality one. Such a phenomenon takes place similarly due to the destructive interference of the two normal modes of the coupled oscillators system \cite{Alzar,Alzar2}. In 2009, Soukoulis and colleagues demonstrated the classical analog of EIT in split ring resonators (SRRs) \cite{Soukoulis2009,Soukoulis2012}. They capacitively coupled a dipole (low-quality) SRR to a quadrupole (high-quality) one by closely placing the gaps of the two SRRs. Electromagnetic drive is on the dipolar oscillator, because incident radiation couples only to dipolar one. They showed that the response of the dipolar SRR exhibit a dip at the resonance frequency which is 43 THz.

Classical analog of EIT is shown to have many applications in the field of plasmonic physics. The coupling of dipole and quadrupole plasmonic modes of two rectangular thin plates (or rectangular gaps in metallic plates) manufactured in sub-micron dimensions are shown to exhibit EIT-like resonances and transparency windows \cite{PlasEIT-Giessen,PlasEIT-Zhang,PlasEIT-Chong,PlasEIT-Kekatpure,PlasEIT-Altug1,PlasEIT-Altug2,YeZhang,PlasEIT-Giessen-2010,PlasmonEIT-absorp,Papasimakis,Gong2012,KivsharRvMdPhys,Longhi2009,ClassEITzheng}. In such devices, slow light propagation \cite{huangveronis,WuShvets}, electromagnetically induced absorption \cite{Soukoulis2012,PlasmonEIT-absorp}, and anomalous light transmission (where covering the subwavelength nanoaperture with gold nanodisk unexpectedly enhances the light transmission) \cite{LightTrans} are observed. Former two effects are proposed to be used in solar cell applications\cite{solarcell2,solarcell}. Because, they trap the light inside solar cell a longer time and provide enhanced light-matter interaction for pair creation \cite{solarcell2}. The life time of the quadrupole moment limits the trapping time in solar cells.

Meanwhile, intense research is being conducted on the control of quantum objects (emitters) by coupling them to plasmonic excitations of nanoantennas \cite{HulstScience}. The plasmon resonance frequencies of nanoantennas are in the optical range and they are tunable by changing the length/width and manufacturing material of the nanoantennas \cite{LivnehNanoLett,Muhlschlegel,FarahaniHecht,nanodiamond,MaksymovArXiv,nanoantennaRvw,Esteban}. In 2010, van Hulst and colleagues coupled a quantum dot (QD) to a gold nanoantenna operating at 800nm \cite{HulstScience}. They showed that QD (excited electronically) transfers the excitation energy \cite{FarahaniHecht} to the nano Uda antenna due to the strong coupling and radiates highly directional. Plasmonic nanoparticles enhance the optical cross-section of a QD five orders of magnitude  by localizing the incident light \cite{QDinducedtrans} This gives rise to potential for radiative communication between optical quantum emitters. It is shown that, such hybrid systems composed of coupled classical and quantum objects also display Fano-like (or EIT-like \cite{PSeit}) resonances \cite{QDinducedtrans,QCfano1,QCfano2,QCfano3,QCfano4,QCfano5,QCfano6,QCfano7,QCfano8,QCfano9}. Absorption spectrum of the plasmonic nanoantenna displays a dip due to the destructive interference of the absorption paths. 

\begin{figure} [htb!]
\includegraphics[width=3.4in]{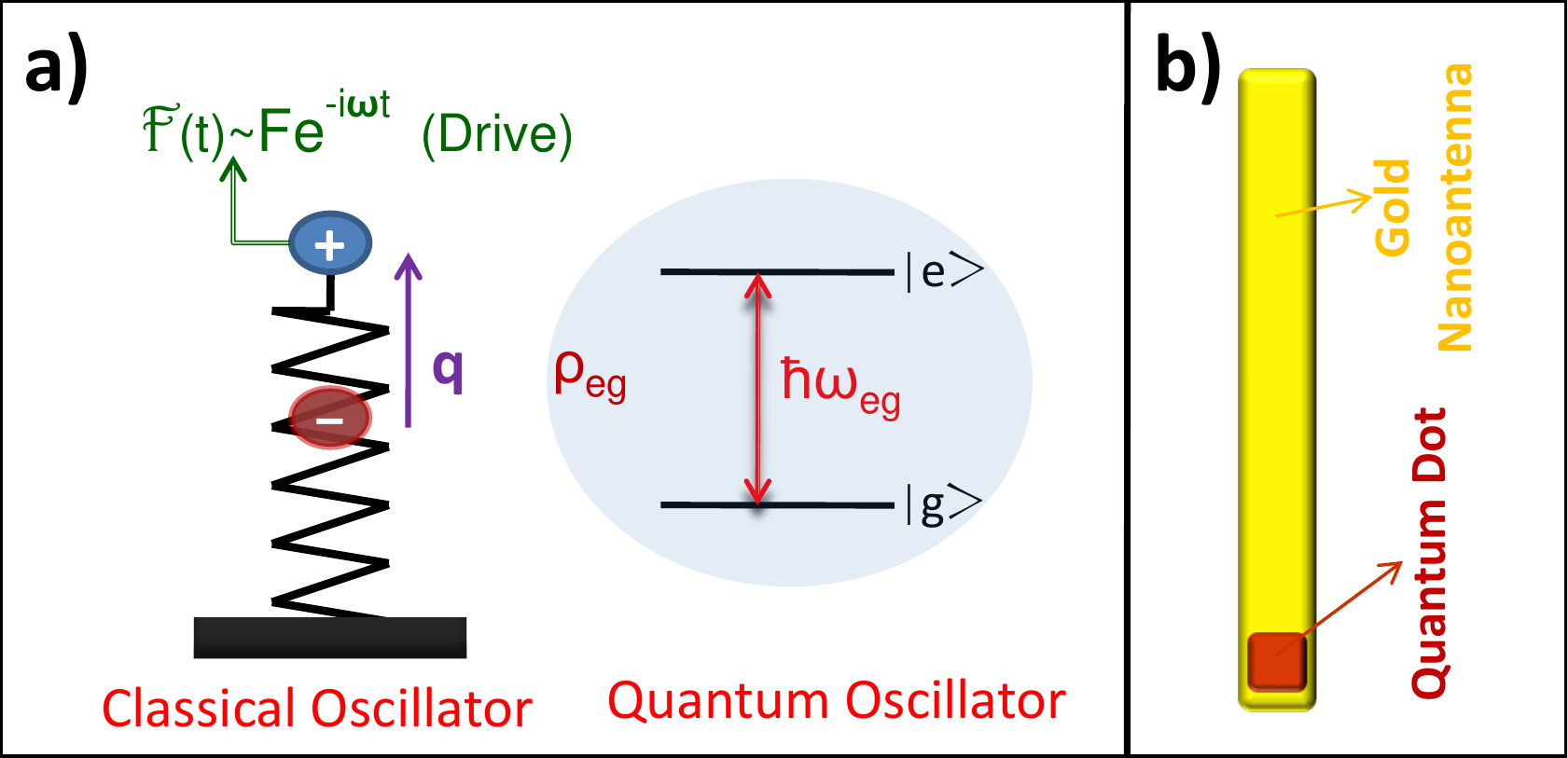}
\caption{(Color online) (a) A classical oscillator driven by an external harmonic source $F(t)=Fe^{-i\omega t}$ is coupled to a quantum oscillator of resonance frequency $\omega_{\rm eg}$. Interaction is dipole-dipole type: The dipole induced on the classical oscillator ($\sim q$) is coupled to the quantum one ($\sim \rho_\text{eg}$). Quantum decay rate ($\gamma_\text{eg}$) is very small compared to damping ($\gamma_q$). (b) The corresponding physical system for the model. A quantum dot is attached over the gold nanoantenna. Plasmonic oscillations on the antenna induces strong electric field at the ends of the bar. The localized electric field ($10^5$ times of the incident one \cite{QDinducedtrans,stockman-review}) couples to the quantum dot. The dipolar excitation of antenna is driven with the incident optical light.}
\label{fig1}
\end{figure}

In this paper, we explore the dynamics of a classical oscillator (plasmonic oscillations on a nanometal) coupled to a quantum emitter (quantum dot \cite{HulstScience,LivnehNanoLett,FarahaniHecht,Pfeiffer2010}, nitrogen-vacancy center \cite{nanodiamond,Zhao2011,Anger2006} or molecule \cite{self-assembled}). We show that, lifetime of the classical oscillator lengthens due to the coupling and it approaches the longer lifetime of the quantum emitter.  The model system is a quantum object attached on the nanoantenna to a position where dipolar mode results very strong electric field localization (Fig.~\ref{fig1}). The localized electric field (five orders larger compared to the incident field \cite{QDinducedtrans}) interacts with the quantum emitter. Hence, a dipole-dipole type effective coupling occurs between the antenna and the quantum object \cite{QDinducedtrans,QCfano1,QCfano2,QCfano3,QCfano4,QCfano5,QCfano6,QCfano7,QCfano8,QCfano9}. Because of the coupling, a transparency window emerges at the center of the absorption spectrum of the antenna, see Fig.~\ref{fig2}. 

The plasmonic oscillations are driven by a harmonic force which stands for the incident light. The coupled system reaches the steady state in much longer times (see Fig.~\ref{fig6}). Thus, nanometal stays polarized for much longer times. Regarding metal nanoparticles, decay rate of the plasmonic excitations ($\gamma_q\sim 10^{14}$Hz , or $\sim 100$meV) \cite{QDinducedtrans,plasmonlife} is very large compared to the decay rate of the quantum objects ($\gamma_{eg}\sim 10^9$Hz, or $\sim 1\mu$eV for QD \cite{QCfano9}). In the coupled system, plasmonic oscillation evolves in time as if it has the tiny decay rate of the quantum emitter. 

This effect has already been shown in the experiment spaser experiment by Noginov {\it et al}. \cite{spaser,spaserPRL,spaserPRL2}. They demonstrated the stimulated emission of plasmonic excitations in a gold nanoparticle core placed in a dye-doped silica shell. They observe a change in the emission spectrum from a broad band to a narrow line ($\sim 10^{12}$Hz, or $\sim 1$meV). The narrowing can be explained within our EIT scheme as follows. The gold nanoparticle core couples to the dye molecules more when silica shell is denser doped. The coupled plasmon--molecule system is governed by the behaviour of the molecular excited level. Thus, the internal conversion rate of the molecule (about $\sim 10^{12}$Hz, or $\sim 1$meV) determines the new plasmon emission band. This is narrower. Additionally, using plasmons with increased lifetime; stimulated emission can dominate the spontaneous emission rate. The spaser establishes. 

The presented effect promises many nanoscale applications. Metal nanoparticle covered solar cell surfaces can trap incident solar light for much longer times. As well, such configurations can lead to slower light propagation. Moreover, increased lifetime enables quantum information processing (including entanglement) with plasmons.

Our simulations also verify a complementary effect. The presence of metal nanoparticle enlarges the decay rate of the quantum emitter \cite{Pfeiffer2010,Zhao2011,Anger2006}, see Fig.~\ref{fig7}.

The paper is organized as follows. In Sec.~\ref{sec:Hamiltonian}, we introduce the Hamiltonian for the coupled classical-quantum oscillator system. We drive the equations of motion for the system using density matrix formalism. We include the damping, quantum decay rates and the drive on the C.O.. In Sec.~\ref{sec:Twindow}, we show the emergence of the EIT-like transparency window. At resonance, power absorption from the drive (by the C.O.) is governed by the quantum decay rate ($\gamma_{\rm eg}$) instead of C.O. damping ($\gamma_{q}$). In Sec.~\ref{sec:analytical}, we adopt an analytical form for the dipolar polarization in the plasmonic mode of the nanoantenna in the vicinity of the resonance. In Sec.~\ref{sec:we-neq-wq}, we picture the shift of the transparency window to the resonance of the quantum emitter ($\omega_{\rm eg}$), when classical and quantum oscillators are not perfectly tuned. In Sec.~\ref{sec:duration}, we show that the durations for reaching the steady-state is governed by the life-time of the quantum object ($1/\gamma_{\rm eg}$) for the coupled system. Thus, plasmons stay polarized for longer times even if the steady-state value for the polarization may vanish. We also discuss the connection with the spaser experiment \cite{spaser,spaserPRL,spaserPRL2} (Sec.~\ref{sec:spaser}) and possible solar cell applications (Sec.~\ref{sec:solar}).  In Sec.~\ref{sec:fluorescence}, we show that life time of the excited quantum emitter is shortened by the plasmon damping. In Sec.~\ref{sec:conclusion}, we summarize the paper and discuss the possible applications of the coupled classical-quantum system.


\section{Analog of EIT in coupled classical-quantum oscillators system}

In this Section, we derive the equations of motion for a system where a classical oscillator (C.O.) is strongly coupled to a quantum emitter. The coupling is through dipole-dipole interaction and resonance frequencies of the oscillators are in the optical regime. We present the effective Hamiltonian for the system and derive the equations of motion using the commutation relations. We plot the power absorbed by the C.O. and depict the emergence of a transparency window (no absorption region) about the resonance frequency.

We consider a system where a quantum object (e.g. quantum dot, molecule or a Nitrogen-vacancy center) is attached over a plasmonic nanoantenna that works in the classical regime \cite{PS1} (see Fig.~\ref{fig1}). The incident light couples to the plasmonic excitation mode of the nanoantenna. Coupling of light to single quantum emitter is of negligible strength compared to the plasmon. The dipole oscillations, which creates a very strong localized electric field over the nanoantenna, couples to the dipolar excitation of the quantum emitter. In Fig.~\ref{fig1}b, in example a quantum dot is placed at a position (end of the annoantenna) where dipolar plasmon mode gives the maximum electric field. Resultant interaction becomes dipole-dipole type. The effective system is: a quantum object coupled to a C.O. which is driven by an external source. The resonance of the plasmonic mode of the C.O. ($\omega_{q}$) is tuned (by varying its size) close to the spacing between the ground and excited levels of the quantum oscillator ($\omega_{\rm eg}$). Here, we consider an oversimplified model of the hybrid system. We mainly aim to demonstrate the phenomenon of lifetime enhancement. In more realistic calculations \cite{QCfano4,QCfano5} one has to consider compicating effects such as the influence of the dielectric environment.

\subsection{Hamiltonian and Equations of Motion} \label{sec:Hamiltonian}

The total hamiltonian ($\hat{H}$) for such a system can be written as the sum of the energy of the quantum object, energy of the plasmonic oscillations, and the dipole-dipole interactions energy
\begin{eqnarray}
\hat{H}_0=\hbar\omega_{\rm e} 
| {\rm e} \rangle \langle {\rm e} | + \hbar\omega_{\rm g}  | {\rm g} \rangle \langle {\rm g}| \: ,
\\
\hat{H}_{q}= \frac{\hat{p}^2}{2m} + \frac{1}{2}m\omega_{q}^2 \hat{q}^2 \: ,
\\
\hat{H}_{\rm int}=\hbar g_c \hat{q} \left( e^{{\rm i}\theta} | {\rm e} \rangle \langle {\rm g} | + e^{-{\rm i}\theta} | {\rm g} \rangle \langle {\rm e}| \right) \: ,
\label{eq:Hint}
\end{eqnarray}
respectively. Dipole moment induced on the antenna (created by the plasmonic oscillations) is proportional to the displacement ($\hat{q}$) of the classical oscillator. The dipole moment of the quantum emitter is proportional to the off-diagonal matrix element $\rho_{\rm eg}$. In Eq. (\ref{eq:Hint}),  $\hat{q}$ (oscillating with freq. $\omega_{q}$)  couples to the quantum dipole excitation ($e^{{\rm i}\theta} |{\rm e}\rangle \langle {\rm g}| + e^{-{\rm i}\theta} |{\rm g}\rangle \langle {\rm e}|$), where $g_c$ is in units of [freq./length] and $e^{{\rm i}\theta}$ is the phase of the matrix element. Parameter $m$ is not directly referred in our derivations, it cancels similar to the case of quantization of Electromagnetic fields (see Chapter. 1 in Ref. \cite{Scullybook}).

We use the commutation relations (e.g. $i\hbar\dot{\hat{q}}=[\hat{q},\hat{H}]$) in deriving the equations of motion (EM). We keep $\hat{q}$ quantum up to a step in order to avoid any fault in the EM. After obtaining the dynamics in the quantum approach, we carry $\hat{q}$ to classical ($q(t)$). Using the rotating wave approximation (RWA), equations take the form  
\begin{eqnarray}
\ddot{q}_0(t)+\gamma_q \dot{q}_0(t)+\omega_q^2 q_0(t)+\frac{\hbar}{m} g_ce^{-{\rm i}\theta} \rho_{\rm eg}^*(t)=\frac{F}{m} e^{-{\rm i}\omega t} \: ,
\label{eq:dq0}
\\
\dot{\rho}_{\rm eg}(t)=({\rm i}\omega_{\rm eg}-\gamma_{\rm eg})\rho_{\rm eg} -{\rm i}g_c e^{-{\rm i}\theta} q_0^* (\rho_{\rm ee}-\rho_{\rm gg}) \: ,
\label{eq:drhoeg}
\\
\dot{\rho}_{\rm ee}(t)=-\gamma_{\rm e}\rho_{\rm ee}(t)-{\rm i} g_c \left( e^{{\rm i}\theta}q_0\rho_{\rm eg} - \text{c.c.}\right) \: ,
\label{eq:drhoee}
\end{eqnarray}
where we use the complex amplitude $q_0(t) \sim e^{-{\rm i}\omega t}$  for describing oscillations. It is related to the displacement of the C.O. as $q(t)=q_0(t)+q_0^*(t)$.  In Eq. (\ref{eq:dq0}), we introduce the harmonic driving force $\left(\frac{F}{m}e^{-{\rm i}\omega t}\right)$, of frequency $\omega$, on the plasmonic oscillator.  Density matrix elements $\rho_{\rm ee}$, $\rho_{\rm gg}$, and $\rho_{\rm eg}$ belongs to the quantum emitter.  $\gamma_{\rm ee}$ is the decay rate of the quantum emitter from the excited state to the ground state. Decay rate of the off-diagonal matrix element $\rho_{\rm eg}$ (or the polarization of the quantum emitter) is represented by $\gamma_{\rm eg}=\gamma_{\rm ee}/2$. The damping rate of the classical (plasmonic) oscillator is $\gamma_q$. Since $\gamma_{\rm ee}$ belongs to a quantum object, we have $\gamma_{\rm ee} \gg \gamma_q$. In example, the typical value for the decay rate of a molecular excitation (determined by internal conversion rate) is $\gamma_{\rm ee}\sim 10^{12}$Hz, whereas the damping rate of plasmonic oscillation is $\gamma_q\sim 10^{14}$Hz$\;$ \cite{plasmonlife}. The constraint on the conservation probability $\rho_{\rm ee}+\rho_{\rm gg}=1$ accompanies Eq.s (\ref{eq:dq0}-\ref{eq:drhoee}).

We seek solutions of the form \cite{Alzar,Alzar2}
\begin{equation}
q_0(t)=\tilde{q}_0e^{-i\omega t} \quad \text{and} \quad \rho_{\rm eg}(t)=\tilde{\rho}_{\rm eg} e^{-{\rm i}\omega t} \: ,
\label{eq:slowvary}
\end{equation}
for the long term (steady-state) behaviour, where $\tilde{q}_0$ is complex number and related to the displacement of the C.O. as
\begin{equation}
q(t)=\tilde{q}_0e^{-i\omega t}+\tilde{q}_0e^{i\omega t}.
\end{equation}
Density matrix element (of amplitude $\tilde{\rho}_{\rm eg}$) oscillates with the source frequency in the steady state.
Inserting Eq. (\ref{eq:slowvary}) into Eq.s (\ref{eq:dq0}-\ref{eq:drhoee}), we obtain the equations
\begin{eqnarray}
(\omega_q^2-\omega^2-{\rm i}\gamma_g\omega) \bar{q} +  f_c \omega_q e^{-{\rm i}\theta} \tilde{\rho}_{\rm eg}^*=\bar{F} \: ,
\label{eq:steady1}
\\
\left[ {\rm i}(\omega-\omega_{\rm eg})+\gamma_{\rm eg} \right]\tilde{\rho}_{\rm eg} + if_ce^{-{\rm i}\theta} \bar{q}^*(\rho_{\rm ee}-\rho_{\rm gg})=0 \: ,
\label{eq:steady2}
\\
\gamma_{\rm ee}\rho_{\rm ee}=-if_c\left( e^{i\theta}\bar{q}\tilde{\rho}_{\rm eg}-e^{-i\theta}\bar{q}^*\tilde{\rho}_{\rm eg}^*\right) \: ,
\label{eq:steady3}
\end{eqnarray}
relating the solutions for the slowly varying variables $\bar{q}$, $\tilde{\rho}_{\rm eg}$ and populations $\rho_{\rm ee}$ and $\rho_{\rm gg}$. We have the additional equation $\rho_{\rm ee}+\rho_{\rm gg}=1$ for the number (probability) conservation. $\bar{q}=\tilde{q}/a_0$ is the dimensionless (scaled) slowly varying (complex) displacement of the C.O., with $a_0=\left(\hbar/m\omega_q\right)^{1/2}$ is the characteristic oscillator length. $f_c=g_ca_0$ is the coupling frequency (strength) and $\bar{F}=F/ma_0$ is the scaled force, in units of $\text{[freq.]}^2$. We note that, in Eq. (\ref{eq:steady2}), scaling results $\frac{\hbar}{m}\frac{g_c}{a_0}=\frac{\hbar}{m}\frac{g_ca_0}{a_0^2}=g_c\omega_q$. Therefore, we do not refer to mass $m$ any more.

\subsection{Transparency Window in Absorption Spectra} \label{sec:Twindow}

We solve the set of nonlinear equations (\ref{eq:steady1}-\ref{eq:steady3}) and we obtain the response of the coupled classical-quantum oscillators system to a driving harmonic field (force). In the solutions, we make the assumption that driving force ($\bar{F}e^{-i\omega t}$) has always been on (till $t=-\infty$), implicitly.

\begin{figure}
\includegraphics[width=3.2in]{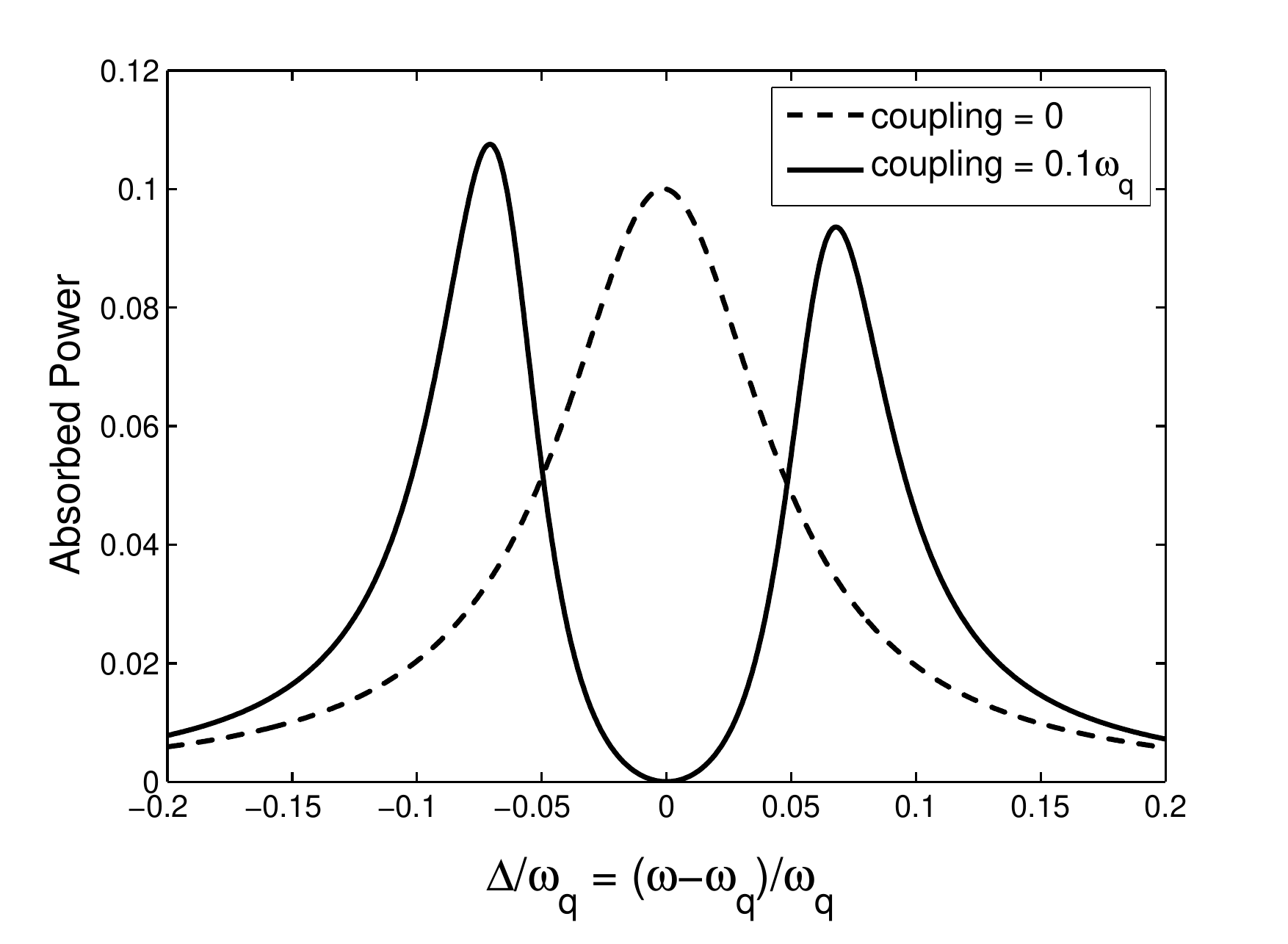}
\caption{The power classical oscillator (nanoantenna) absorbs from the driving force (electromagnetic field). When the nanoantenna operates alone, absorption is peaked (dashed-line) about the resonance frequency frequency of the antenna ($\omega_q$). If the nanoantenna is coupled to a quantum oscillator (e.g. quantum dot), a transparency window occurs (solid-line) in the mid of the absorption peak. The new absorption at resonance is determined by the decay rate of the quantum dot ($\gamma_\text{eg} \sim 10^9$Hz), which is several orders of magnitude smaller compared to the damping ($\gamma_q\sim 10^{14}$Hz). The quantum level spacing is perfectly tuned with the resonance of antenna, e.g. $\omega_\text{eg}=\omega_q$. The damping and quantum decay rates $\gamma_q=0.1\omega_q$ and $\gamma_\text{eg}=0.5\times 10^{-4}\omega_q$ are used in the simulation. Scaled driving force is $\bar{F}=0.01$ (in units of $\omega_q^2$).}
\label{fig2}
\end{figure}

Regarding the plasmonic oscillations, $q(t)=\tilde{q}_0e^{-i\omega t}+\tilde{q}_0e^{i\omega t}$ corresponds to dipolar polarization on the antenna. Thus, real/imaginary part of $\tilde{q}_0$ refers to the polarization/absorption. This is analogues to the relation  ${\bf P_{\omega}}=\chi(\omega){\bf E_{\omega}}$, where ${\bf P}$, ${\bf E}$, and $\chi$  stand for Polarization, Electric field, and dielectric susceptibility, respectively. An equivalent relation can be calculated also from the power that force does on the oscillator (force$\times$velocity) \cite{Alzar,Alzar2} per cycle,
\begin{equation}
P(t)={\rm Re} \left\{-i\omega F \tilde{q} \right\}={\rm Im} \left\{ \omega F \tilde{q}\right\} \: .
\end{equation}
Therefore, absorbed power is proportional to the imaginary part of the $\tilde{q}$, as discussed above. In Fig.~\ref{fig2}, we plot the absorbed power for varying drive (source) frequency $\omega$. We take the resonance frequency of the plasmonic mode equal to the quantum level spacing, i.e. $\omega_\text{eg}=\omega_q$. When the C.O. is not coupled with the quantum object, absorption shows a peak at the resonance $\omega=\omega_q$ (dashed-line in Fig.~\ref{fig2}). On the other hand, if the C.O. is coupled to the Q.O. with coupling strength $f_c=0.1\omega_q$, emergence of a transparency window at the center of the absorption peak ($\omega=\omega_q$) is observed (solid-line in Fig.~\ref{fig2}). In the vicinity of $\omega=\omega_q$, the absorbance is proportional to the quantum decay rate $\gamma_\text{eg}$ instead of the classical one $\gamma_\text{q}$ similar to Ref.s \cite{Scullybook,Alzar,Alzar2} (see Eq. (\ref{eq:Pow-resonance}) below). The driving force is $\bar{F}=0.01$ in units of $\omega_q^2$, and corresponds to an pump intensity of about $10^3$mW/$\text{cm}^2$. 

\begin{figure}
\includegraphics[width=3.2in]{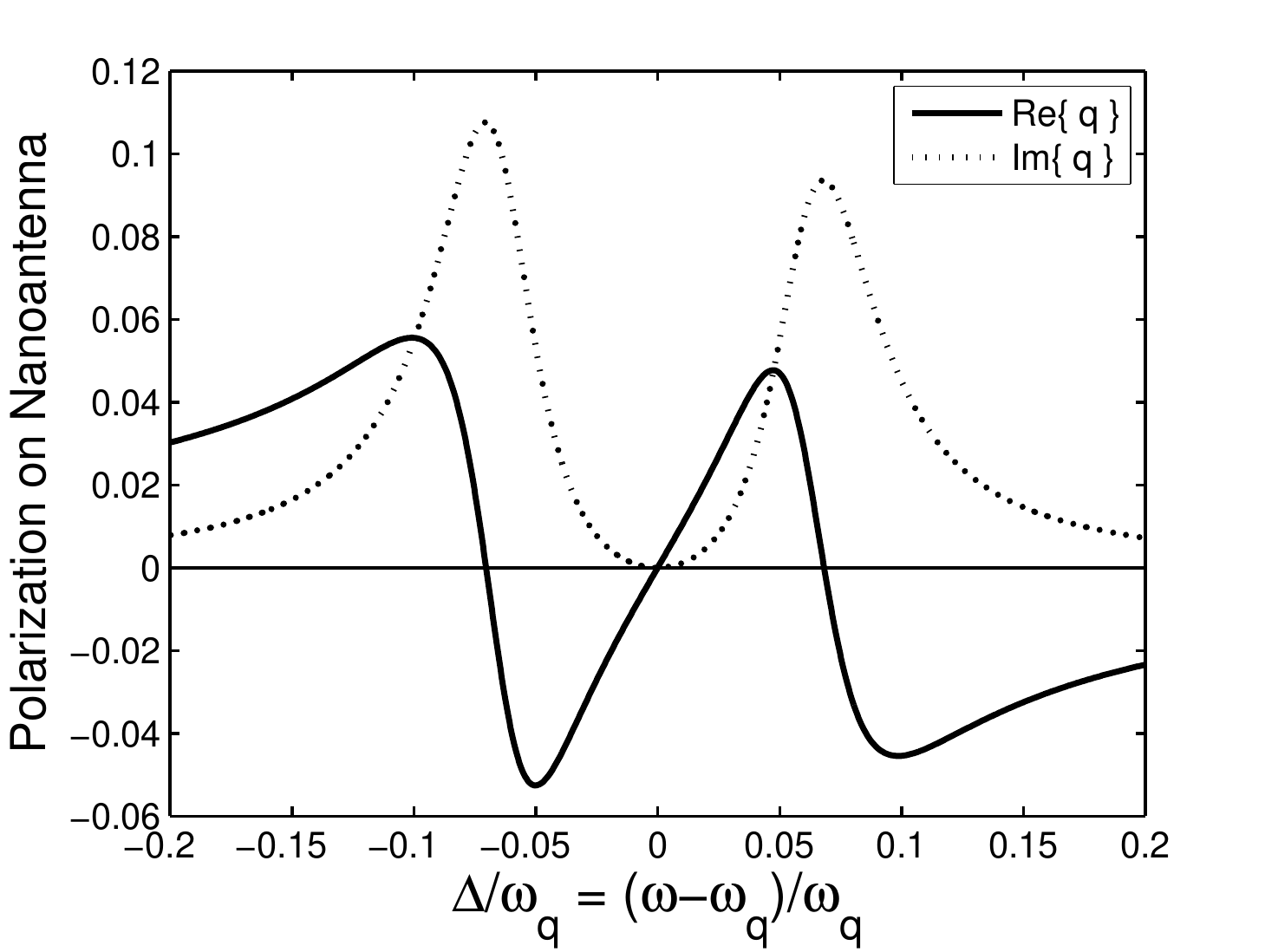}
\caption{The real (solid-line) and the imaginary (dotted-line) parts of the displacement ($\bar{q}$) of the classical oscillator. Real part corresponds to the polarization induced on the nanoantenna. Imaginary part is proportional to the antenna absorption. We clearly observe the EIT-like behaviour \cite{Scullybook,Alzar,Alzar2,Soukoulis2009,Soukoulis2012}.  The parameters are the same with Fig.~\ref{fig2}.}
\label{fig3}
\end{figure}

Regarding plasmonic oscillations, $q(t)$ corresponds to polarization field in the antenna. So, In Fig.~\ref{fig3}, we plot the nanoantenna polarization ${\rm Re}\{\bar{q}\}$ (solid-line) together with the antenna absorption (${\rm Im} \{\bar{q}\}$). Fig.~\ref{fig3} depicts the common form of the EIT-like response \cite{Scullybook}. 

\begin{figure}
\includegraphics[width=3.2in]{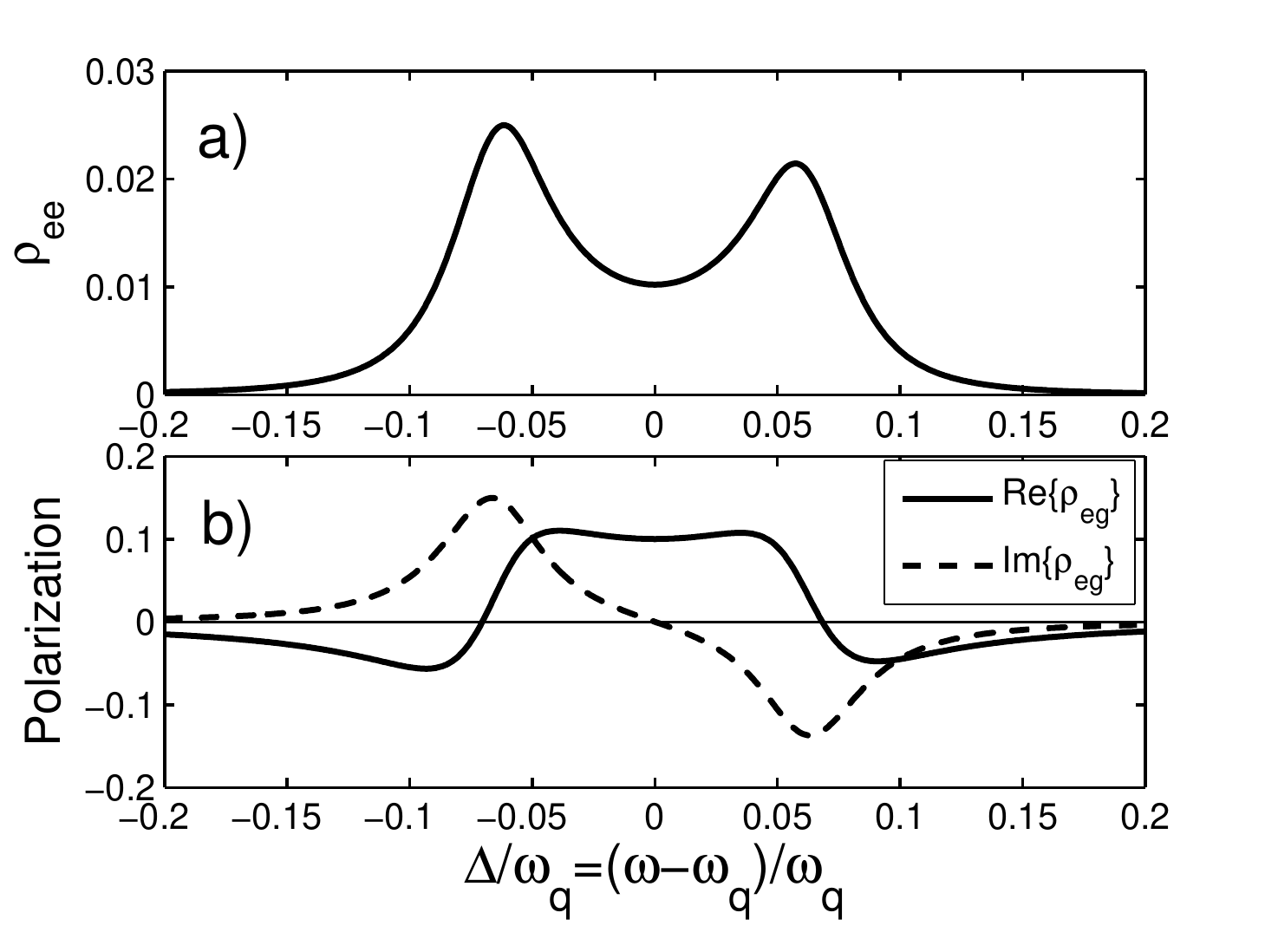}
\caption{ The corresponding steady state values for the (a) excitation fraction and (b) polarization in the quantum emitter, which is coupled to the nanoantenna (classical oscillator). At resonance, quantum emitter stays polarized (${\rm Re}\{\rho_\text{eg}\}\neq 0$) with a finite excitation probability ($\rho_\text{ee} \neq 0$). The absorption of the quantum object vanishes at resonance (${\rm Im} \{\rho_\text{eg}\}=0$). }
\label{fig4}
\end{figure}

Fig.~\ref{fig4} plots the corresponding values for the polarization in the quantum dot ($\tilde{\rho}_\text{eg}(\omega)$) and the excitation fraction ($\rho_\text{ee}(\omega)$). We see that (solid-line in Fig.~\ref{fig4}b), quantum dot stays polarized (${\rm Re}\left\{\rho_\text{eg}\right\} \neq 0$) without absorption (${\rm Im}\left\{\rho_\text{eg}\right\} = 0$) at resonant drive $\omega=\omega_q$. In addition, quantum dot becomes partially excited ($\rho_\text{ee} \neq 0$) about the resonance. The reason for the asymmetric absorption peaks in Fig.~\ref{fig2} is due to the antisymmetric absorption profile (about $\Delta=0$) of the quantum oscillator. For $\omega > \omega_q$, quantum object display gain (${\rm Im}\left\{\rho_\text{eg}\right\} < 0$) in the coupled system. Such asymmetric absorption (scattering) profiles are common to Fano resonances \cite{KivsharRvMdPhys}.
%
%

\subsection{Analytical form for antenna polarization ($\bar{q}$)} \label{sec:analytical}

Exact expressions for $\bar{q}$ are cumbersome since they include the solutions of the cubic equations. However, analytical expressions where the solution for the population inversion $y=\rho_\text{ee}-\rho_\text{gg}$ left implicit give us clues about the response of the coupled system. When the expression for $\rho_\text{eg}$, obtained from Eq. (\ref{eq:steady1}), is used in Eq. (\ref{eq:steady2}), one obtains
\begin{equation}
\bar{q}=\bar{F} \frac{(\omega-\omega_\text{eg})+i\gamma_\text{eg}}{D_{\rm R}+iD_{\rm I}} \: ,
\label{eq:q}
\end{equation}
where 
\begin{eqnarray}
D_{\rm R}=(\omega-\omega_\text{eg})(\omega_q^2-\omega^2)+\gamma_q\gamma_\text{eg}\omega -f_c^2\omega_q y
\\
D_{\rm I}=\gamma_\text{eg}(\omega_q^2-\omega^2)-\gamma_q\omega(\omega-\omega_\text{eg})
\end{eqnarray}
are the real and imaginary parts of the denominator in Eq. (\ref{eq:q}). $\bar{F}$ is the scaled force in units of $\text{[freq.]}^2$. The value of population inversion $y$ is determined by solving the cubic equation
\begin{equation}
\frac{1+y}{2}=-\frac{f_c^2y}{(\omega-\omega_\text{eg})+\gamma_\text{eg}^2}|\bar{q}|^2 \: ,
\label{eq:yeq}
\end{equation}
which is obtained by using the expression for $\rho_\text{eg}$ in Eq. (\ref{eq:steady3}). From Eq. (\ref{eq:yeq}) it can be seen that $y<0$ must be satisfied. Among the three roots for $y$, only one is real or satisfies all of the equation (\ref{eq:steady1}-\ref{eq:steady3}) \cite{PS2}.

When \underline{$\omega=\omega_q=\omega_\text{eg}$}, Eq. (\ref{eq:q}) simplifies to
\begin{equation}
\bar{q}=i\frac{\bar{F}\gamma_\text{eg}}{\left(\gamma_q\gamma_\text{eg}-f_c^2y\right)\omega_q} \: ,
\label{eq:17}
\end{equation}
that is purely imaginary and resulting no polarization. Since $\gamma_\text{eg}$ is very small compared to all frequencies, absorption is represented by
\begin{equation}
P\simeq \bar{F} \gamma_\text{eg} / (-f_c^2y)
\label{eq:Pow-resonance}
\end{equation}
which is proportional to $\gamma_\text{eg}$, not $\gamma_q$ \cite{Alzar,Alzar2}. This simple result depicts the expected absorption cancellation effect common to EIT response. For $\omega=\omega_q=\omega_\text{eg}$ and $\gamma_\text{eg} \rightarrow 0$, the equation for $y$ [that is Eq. (\ref{eq:yeq})] simplifies to
\begin{equation}
y^2+y+c=0 \; ,
\label{eq:yeqsimlified}
\end{equation}
where $c=2\bar{F}^2/f_c^2\omega_q^2$. Eq.(\ref{eq:yeqsimlified}) has real solutions only if discriminant is nonnegative, that is if $\bar{F} < \left(f_c\omega_q/2\sqrt{2}\right)$. Hence, in the linear regime (when drive is small) $y$ has the real solutions 
\begin{equation}
y_{1,2} \simeq -\frac{1}{2} \pm \frac{(1-4c)^{1/2}}{2} \: .
\label{eq:20}
\end{equation}

In our simulations, we verify that Eq.s (\ref{eq:17},\ref{eq:Pow-resonance},\ref{eq:20}) describe the absorption in the steady state --for resonantly matched classical, quantum oscillators with the source-- provided that $\gamma_{eg}$ is small compared to other frequencies.

\subsection{Shift of Trancparency Window when C.O. is not perfectly tuned to Q.O. ($\mathbf{\omega_{\rm eg}\neq \omega_{\rm q}}$)} \label{sec:we-neq-wq}

\begin{figure}
\includegraphics[width=3.2in]{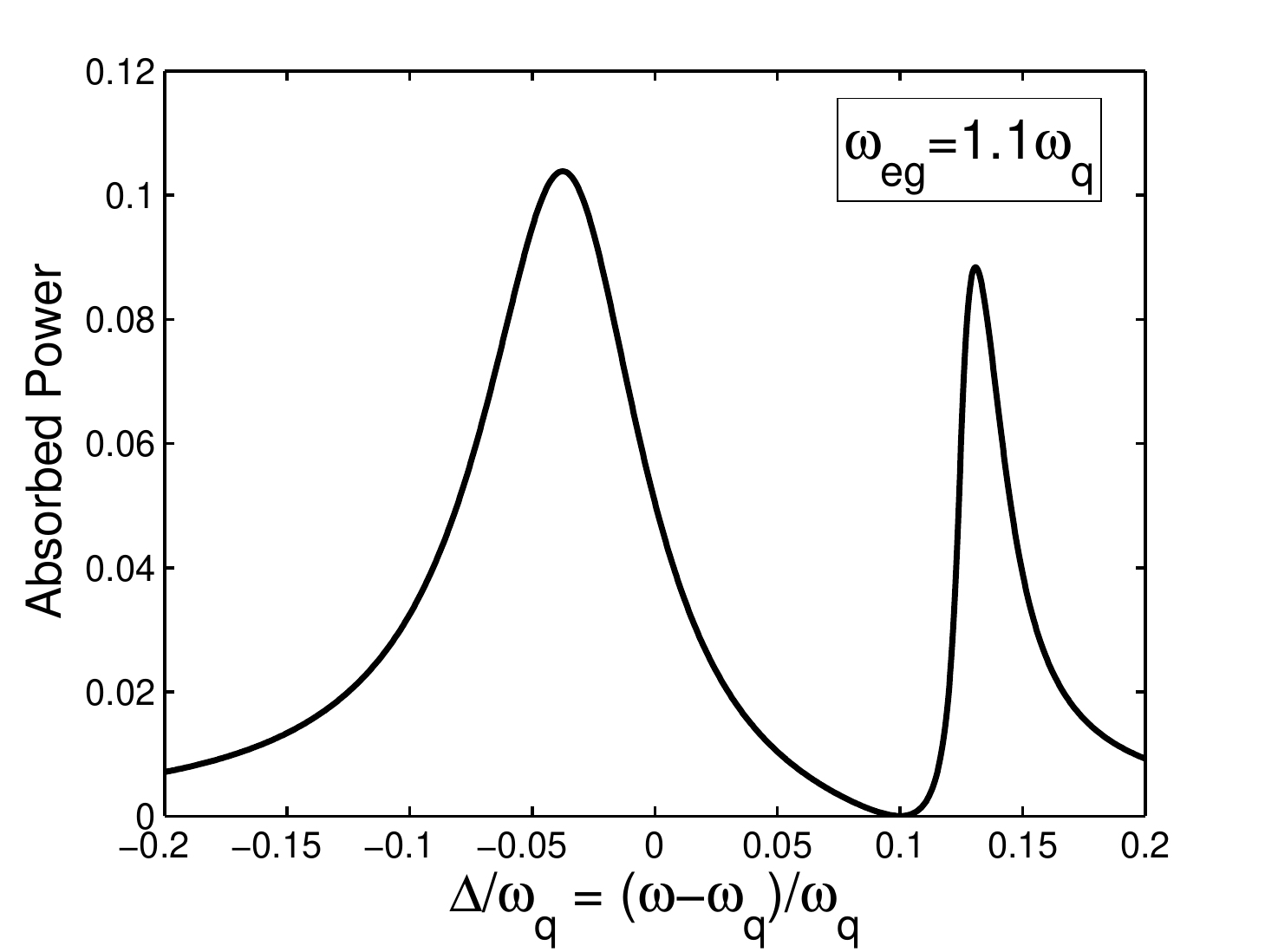}
\caption{The spectral variation of the absorbed power on the nanoantenna when the resonance frequency of the antenna ($\omega_q$) is not perfectly tuned to the quantum level spacing ($\omega_\text{eg}$). The position, where EIT-like response occurs, shifts to the quantum resonance $\omega=\omega_\text{eg}=1.1\omega_q$. Antenna can rearrange its operation frequency with in its spectral width $\pm\gamma_q=\pm0.1\omega_q$ in order to match the quantum oscillation. The coupling strength is $f_c=0.1\omega_q$ and scaled driving force is $\bar{F}=0.01$ (in units of $\omega_q^2$). }
\label{fig5}
\end{figure}

When the resonance frequency of the dipole mode of the plasmonic excitation of the nanoantenna does not match the quantum level separation ($\omega_\text{eg}$), the bahavior of the EIT-like response is modified. The position, where EIT occurs shifts to $\omega \simeq \omega_\text{eg}$ (see Fig.~\ref{fig5}). This takes place due to the following effect. The spectral width of the quantum oscillator (quantum dot) $\Delta_\text{qua} \simeq \gamma_\text{eg}=0.001\omega_q$ is very tight compared to the spectral width of the C.O. $\Delta_\text{cls}=0.1\gamma_q$. Thus, in order to establish the coupling, C.O. rearranges its frequency (within $\Delta_q \sim \gamma_q$) in order to match the energy spacing of the quantum system.
%
%

\section{Enhaced Duration for Plasmonic Oscillations} \label{sec:duration}

In Fig.~\ref{fig6}, we plot time evolutions of the antenna polarization ($\bar{q}={\rm Re}\{\bar{q}\}+i{\rm Im}\{\bar{q}\}$) when there is no coupling (Fig.~\ref{fig6}a) between the classical-quantum oscillators and in the existence of coupling $f_c=0.1\omega_q$ (Fig.~\ref{fig6}b). Quantum level separation is $\omega_\text{eg}=1.01\omega_q$, decay rates are $\gamma_q=0.1\omega_q$ and $\gamma_\text{eg}=0.0005\omega_q$. We observe that coupled oscillator reaches the steady state in the order of $\gamma_q/\gamma_\text{eg} \sim 200$ times longer compared to the uncoupled one. Thus, the coupled system remains polarized for a much longer time even if the final polarization for both case may be zero at the steady state (not zero in Fig.~\ref{fig6}b).

In this paper, we aim to present simply the possibility of the lifetime enhancement for a C.O. (plasmonic field) without concentrating on the specific system parameters. Even though the typical frequency values for nanoantenna plasmons can be specified, the damping rates for quantum emitters occupy a wide range. The surface plasmon frequency for metal nanoparticles are in the optical range $\omega_q \sim  10^{15}$Hz (or $\sim 1$eV). The typical decay rate for plasmons is $\gamma_q\sim 10^{14}$Hz$\;$ \cite{QDinducedtrans,plasmonlife}.  On the other hand, the decay rate for quantum objects varies. An Alkali atom can have $\gamma_q \sim 10^7$Hz. The typical value for QD is about $\gamma_q \sim 10^{9}$Hz $\;$ \cite{QCfano9}. In dye molecules, even though the radiative lifetime is as long as nanoseconds, the internal conversion rate ($\gamma_q\sim 10^{12}$Hz) determines the duration of the excitation.

\subsection{The Spaser Experiment} \label{sec:spaser}

\begin{figure}
\includegraphics[width=3.2in]{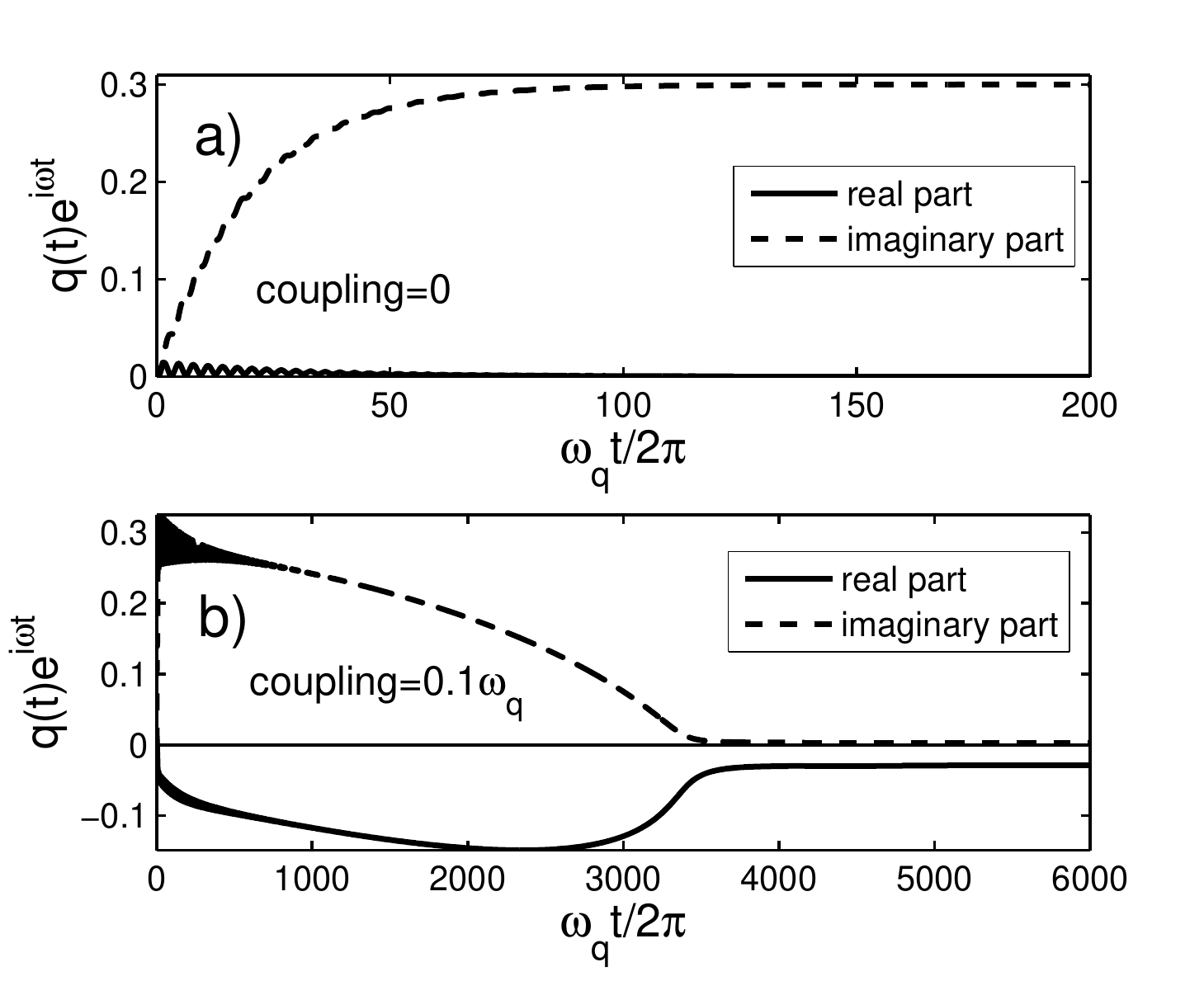}
\caption{Comparison of the durations for the two system to reach the steady-state polarization. The C.O. is driven by the harmonic force $F(t)=Fe^{-i\omega t}$. (a) Nanoantenna (C.O.) is not coupled to the quantum emitter (e.g. a molecule). Antenna polarization (${\rm Re}\{q\}$) reaches the steady-state at about $\omega_qt=2\pi \times 20$. (b) Nanoantenna is coupled to the quantum dot with coupling strength $f_c=0.1\omega_q$. Antenna reaches the steady-state in about $\omega_qt=2\pi \times 4000$. Coupled nanoantenna stay polarized for about $\gamma_q/\gamma_\text{eg} \simeq 200$ times longer compared to the uncoupled one.  Lifetime enhancement in plasmon excitations enables the use of metal nanoparticles in plasmonic quantum information and in solar cell applications \cite{solarcell2,solarcell,dewetting,gursoy}. Quantum level spacing is $\omega_\text{eg}=1.01\omega_q$. Decay rates are $\gamma_q=0.1\omega_q$ and $\gamma_\text{eg}=0.0005\omega_q$. Driving force is $\bar{F}=0.03$ (in units of $\omega_q^2$) and drive frequency is $\omega=\omega_q$. Parameters correspond to spaser experiment \cite{spaser,spaserPRL,spaserPRL2}. }
\label{fig6}
\end{figure}

The scaled values in Fig.~\ref{fig6} corresponds to a plasmon-molecule coupled system with frequencies $\omega_q\sim 10^{15}$Hz, $\gamma_q\sim  10^{14}$Hz and $\gamma_{ee}\sim 10^{12}$Hz ($\gamma_{eg}=\gamma_{ee}/2$). The coupling frequency $f_c=\gamma_q$ may be varied by closer placing the molecule-nanoparticle system. Such strong coupling regimes ($f_c > \gamma_q$) are achievable \cite{fret}. This system has been experimented by Noginov {\it et al.} \cite{spaser,spaserPRL,spaserPRL2} within the spaser (surface plasmon amplification by stimulated emission of radiation) concept. For stronger coupling between molecule-doped silica shell and gold nanoparticle core --by increasing the doping concentration-- a narrower emission band for the combined system system can be observed. Decreasing the spontaneous emission rate (plasmon decay rate), above a critical pump stimulated emission of plasmons wins against spontaneous one.

It is natural to expect the emergence of the lifetime enhancement effect after a sufficient driving force (pump intensity). The induced dipole field must be high enough for coupling energy to be in a similar order with the plasmon decay rate. Because, energy transfer rate from plasmon excitation to the quantum emitter must be not negligible compared to the energy transfer rate to the vacuum modes.

Noginov {\it et al.} points out that; when the density of the hybrid nanoparticles is decreased (not the damping concentration for a single nanoparticle), the emission intensity decreases but the narrowing in the emission spectrum is preserved. This shows that, spectral narrowing originates from a single hybrid system a critical doping (molecule) concentration. Thus, lasing originates not from the collective act of the ensemble of the hybrid systems. It rather originates from the extension of the plasmon lifetime in a single system due to the stronger molecule-plasmon coupling.

In order to observe the duration enhancement for the plasmonic oscillation (polarization) one does not need a perfect matching between the plasmon resonance frequency and the quantum level spacing. In Fig.~\ref{fig5}, we observe that position of the EIT center is determined by the quantum emitter.

A similar lifetime enhancement effect occurs when a dipolar plasmon mode is coupled to a high-life quadrupole plasmon mode: full classical analog of EIT \cite{solarcell2,solarcell,dewetting,gursoy}. However, the lifetime for a quantum excitation is much longer than the classical quadrupole plasmon one.

\subsection{Nanoscale Applications} \label{sec:solar}

The common aim in the solar cell research is to increase the time that sun light spends in/on the semiconductor panels \cite{solarcell2,solarcell,gursoy,dewetting,huangveronis,WuShvets}. The incident sun light is trapped in the silver/gold nanospheres (of diameter ~80nm) which are placed on the surface of the solar cells by dewetting technique \cite{solarcell2,dewetting,gursoy}.

In a coupled metal nanosphere--quantum emitter system, light trapping duration can be extended upto $10^{-9}$s (for certain drive frequencies)  using quantum dots (QDs). This increases the pair formation efficiency in solar panels. In Ref.~\cite{self-assembled}, self-assembled manufacturing of bio-molecules that are coupled to gold nanospheres reported. Random distributions of such objects on the solar cell surface is possible to work as EIT centers. Alternatively, sputtering molecules over the surface can be expected to work. Using molecules with different quantum level spacings, EIT centers can cover different frequency ranges. A variety of molecules can be fit into a wavelength square. Even though QDs display higher lifetime, they are not appropriate for self-assembled (mass production) techniques. 

On the other hand, QDs can be used in quantum information processing. Plasmons (of high electromagnetic cross section) with increased lifetime can be utilized in processes which need quantum entanglement.

\subsection{Shortened decay time for quantum emitter } \label{sec:fluorescence}

\begin{figure}
\includegraphics[width=3.2in]{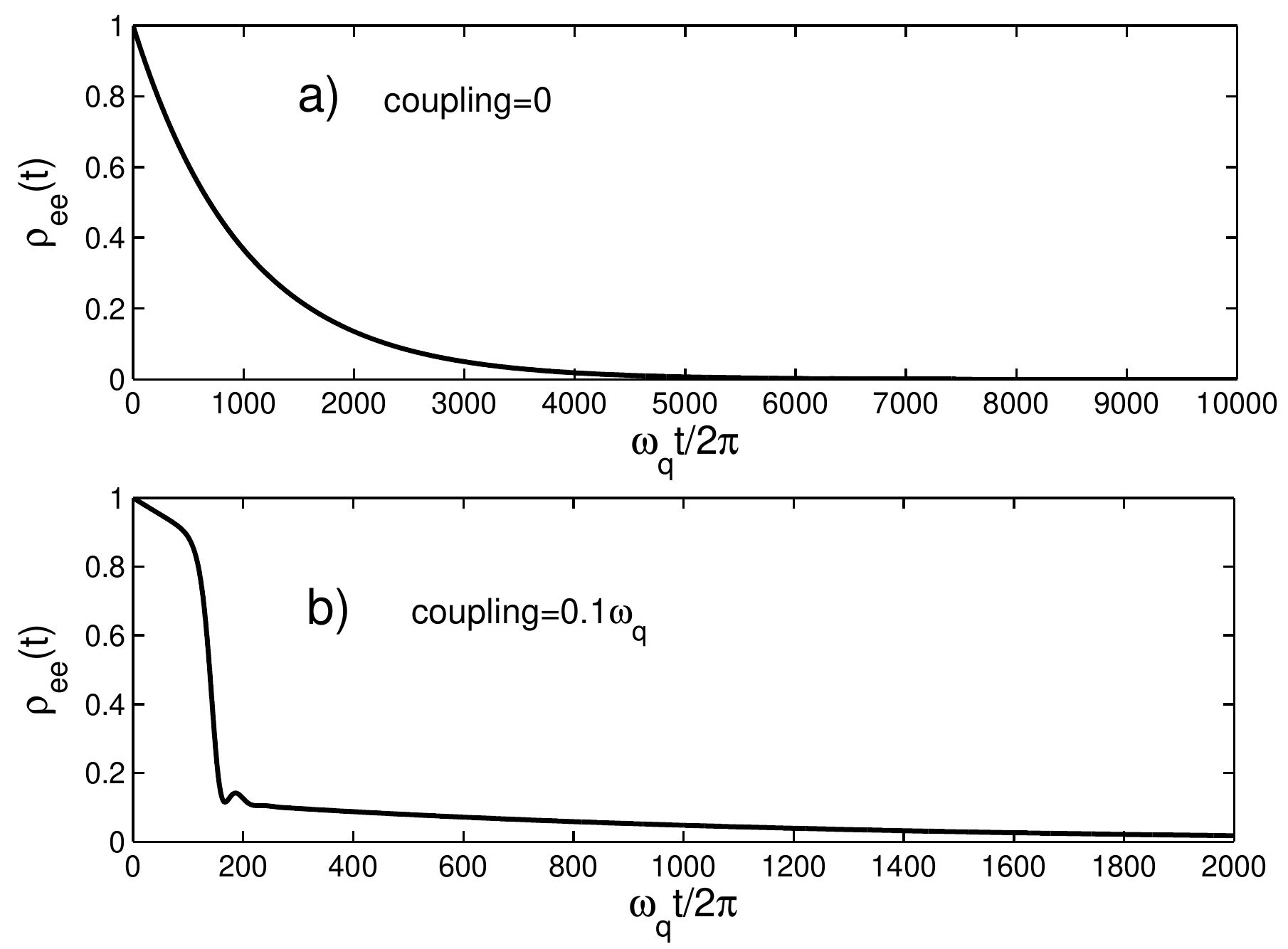}
\caption{Comparison of the durations for the excited quantum emitter to decay into the ground state. Quantum emitter is prepared initially in the excited state ($\rho_{\rm ee}=1$) and there is no applied drive. (a) Quantum dot is not coupled to nanometal. Decay to ground state happens in natural decay time $1/\gamma_{\rm eg} \sim 2000$. (b) Quantum dot is coupled to metal nanoantenna with strength $f_c=0.1\omega_q$. The decay of the composite system takes only $\sim$ 200$/\omega_q$ time. Quantum level spacing is $\omega_\text{eg}=1.01\omega_q$. Decay rates are $\gamma_q=0.1\omega_q$ and $\gamma_\text{eg}=0.0005\omega_q$.}
\label{fig7}
\end{figure}

We additionally verify a complementary  effect: decay time of the electronically excited quantum emitter to the ground state is shortened due to the presence of the metal nanoparticle \cite{Pfeiffer2010,Zhao2011,Anger2006}. In Fig.~\ref{fig7}, the quantum dot is prepared initially in the excited state ($\rho_{\rm ee}=1$). There exists no drive, but quantum dot is coupled with a strength of $f_c=0.1\omega_q$ to the nanoantenna. Comparing Fig.~\ref{fig7}a and Fig.~\ref{fig7}b, the coupled composite system emits the energy in a much shorter time (about 10 times).

The increased quantum decay rate can also be attributed to the change in the density of states due to the hybridization of the quantum dot and the nanometal \cite{Anger2006} (Purcell effect). Here, we observe a parallel effect due to the modification of the effective decay rate of the quantum emitter by the damping rate of the plasmonic antenna. This effect has totally different origins. Therefore, the observed shorting has dual origins both working in the same direction (shortens).


\section{Summary and Conclusions} \label{sec:conclusion}

Incident sun light couples very strongly to the metal nanoparticle plasmons. The trapped intense radiation results enhanced light-matters interaction \cite{solarcell2}. However, this oscillations last in only about $1/\gamma_q \sim 10^{-14}$s and light is re-radiated in different frequencies (heat). We show that the lifetime of the light trapping in plasmonic oscillations can be increased upto the quantum decay time $1/\gamma_{\rm eg} \sim 10^{-9}$s. The dynamics of the driven plasmons is governed by the lifetime of the quantum emitter (see Fig.~\ref{fig6}) which is coupled to it. The new lifetime is limited with the coherence time of the driving source \cite{PS3,mandelwolf}.

The effect is observed in the experiment \cite{spaser,spaserPRL,spaserPRL2} demonstrating the stimulated emission of plasmon excitations (spaser). The emission band of the coupled metal nanoparticle--molecule system is determined by the internal conversion rate of the dye molecule.

The presence of such an effect is due to the semiquantum--semiclassical analog of electromagnetically induced transparency (EIT) \cite{Scullybook}. The coupling of the quantum object creates an additional oscillation mode whose frequency is within the uncertainty window (proportional to the damping rate) of the classical oscillator. The two normal modes of the classical oscillator interfere destructively and absorption cancels.  A transparency window emerges at the center of the absorption peak of the plasmonic nanometal. A similar version of duration enhancement also occurs in the full-classical analog of EIT \cite{Alzar,Alzar2,Soukoulis2009}, where a dipole mode is coupled to a high-life quadrupole mode  \cite{Soukoulis2012,dewetting,gursoy}. However, the lifetime of quantum excitation is highly large compared to the quadrupole mode plasmon lifetime.

The EIT in coupled classical--quantum systems enables us to benefit from useful features of both systems. In such a devise, one can combine large interaction cross section of plasmons (with light) and long decay time of the quantum object. This allows one to use nanoantennas in quantum information. Such an effect can also be adopted in solar cell applications for longer trapping of light.

In addition, the slow light propagation occurring within the transparency window last much longer. Regarding the macroscopic light propagation inside the solar cell, this effect additionally increases the optical path length.

\begin{acknowledgements}
I acknowledge  support  from T\"{U}B\.{I}TAK-KAR\.{I}YER  Grant
No.  112T927 and T\"UB\.{I}TAK-1001 Grant No. 110T876. I specially thank G\"{u}rsoy B. Akg\"{u}\c{c} for guiding discussions and his intensive help with the manuscript. I specially thank  T. \c{C}etin  Ak{\i}nc{\i} and
Serhat \c{S}eker for their motivational support.
\end{acknowledgements}


%
%
%

\begin{thebibliography}{99}
\bibitem{PS3} The phase of the sunlight exhibit random fluctuations. This effect has second quantized origin (see Chapter 14.5 in Ref.~\cite{Scullybook} and Ref.s~\cite{BECsqueezed1,BECsqueezed2}) and gives rise to a finite coherence time ($\tau\sim 10^{-8}$s) for the incident sun light. For this reason, at the end of each phase-coherence period, one shall consider a reset source. The shift in the oscillation phase of the source can interfere destructively with the previous polarization constructed on the nanometal. Thus, plasmonic and quantum oscillations are reorganized after each phase-coherence period.
\bibitem{PS2} Near the resonance $\omega=\omega_q=\omega_\text{eg}$, there occurs a second real root for $y$. The system of equations we study is nonlinear. Thus, convergence to one of the steady states depends on the initial value \cite{chaosbook}. We check that; for the initial conditions $\rho_\text{gg}(t=0)=1$ , $\rho_\text{eg}(t=0)=0$ and $q(t=0)=0$, real-time simulations of Eq. (\ref{eq:dq0}-\ref{eq:drhoee}) converges to the plotted results.
\end{thebibliography}
\end{document}